\documentclass[article,showpacs,nofootinbib]{revtex4-1}

\usepackage{amsmath}
\usepackage{graphicx}
\usepackage{epsf}
\usepackage{psfrag}
\usepackage{epsfig}
\usepackage{graphics}
\usepackage{amsfonts}
\usepackage{epstopdf}
\usepackage{slashed}
\usepackage{footnote}

\newcommand{\be}{\begin{equation}}
\newcommand{\ee}{\end{equation}}
\newcommand{\bq}{\begin{eqnarray}}
\newcommand{\eq}{\end{eqnarray}}
\newcommand{\ba}{\begin{align}}
\newcommand{\ea}{\end{align}}

\begin{document}

\title{Momentum Routing Invariance in Extended QED: Assuring Gauge Invariance Beyond Tree Level}

\date{\today}

\author{A. R. Vieira$^{(a,b)}$} \email[]{arvieira@indiana.edu}
\author{A. L. Cherchiglia$^{(c)}$} \email[]{adriano.cherchiglia@mailbox.tu-dresden.de}
\author{Marcos Sampaio$^{(a)}$} \email []{msampaio@fisica.ufmg.br}

\affiliation{(a) Departamento de F\'isica - ICEx - Universidade Federal de Minas Gerais\\ P.O. BOX 702, 30.161-970, Belo Horizonte - MG - Brasil}
\affiliation{(b) Indiana University Center for Spacetime Symmetries, Bloomington, Indiana 47405, USA}
\affiliation{(c) Institut fur Kern- und Teilchenphysik, TU Dresden, Zellescher Weg 19, 01069 Dresden, Germany}

\begin{abstract}
We address the study of gauge invariance in the Standard Model Extension which encompasses all Lorentz-violating terms originated by spontaneous symmetry 
breaking at the Planck scale. In particular, we fully evaluate Ward identities involving two and three point functions and derive the conditions which assure 
gauge invariance of the electromagnetic sector of the Standard Model Extension at one-loop. We show that  momentum routing invariance is sufficient to fix 
arbitrary and regularization dependent parameters intrinsic to perturbation theory in the diagrams involved. A scheme which judiciously collects finite but 
undetermined quantum corrections is employed, a particularly subtle issue in the presence of $\gamma_5$ matrices.
\end{abstract}


\maketitle

\section{Introduction}

The Poincar\'e group determines to a large extent the general structure of all relativistic quantum field theories. Enlarging its symmetry content to conformal and super-Poincar\'e groups enabled  theoreticians to speculate about physics beyond the standard model. In other words, physical theories are essentially determined by their invariances (symmetries),  the form of interactions being the preeminent example which results from the interplay between Lorentz covariance and gauge symmetry. However it is still necessary to specify at which scale (macroscopic or microscopic, low or high energy, etc.), at which level (classical or quantal) and what type (exact, approximate or broken) the symmetry content of a model effectively describes certain physical phenomena. Symmetry breaking in physics are usually explicit, anomalous or spontaneous. In group theoretical terms an unbroken larger group is broken to its subgroups.

Charge conjugation (C), time reversal (T) and parity (P) transformations are independently respected by strong and electromagnetic interactions whereas P 
violation is a signature of weak interactions. Baryon asymmetry in the universe relies strongly  on CP violation which motivates the search of additional 
mechanisms beyond those observed and predicted by the standard model. However the discrete symmetries C, P and T are related by the CPT theorem, demonstrated by 
Luders in 1952, which states that the combination of C, P, and T is a general symmetry of local relativistic field theories without gravity. CPT symmetry can be 
broken as a result of Lorentz violation \cite{Samuel} as seen in the Standard Model Extension (SME) \cite{Colladay}, the low energy framework  for 
Lorentz-breaking effects. The SME is built with all Lorentz and CPT violating lagrangian terms consistent with coordinate invariance \cite{Lehnert}. It 
allows for the identification and comparison of most currently feasible experiments searching for Lorentz
and CPT violation. Any signal of a tiny coefficient of a Lorentz-violating term would support the idea of an unified theory at Planck scale. 
With that purpose, several tests have been performed with the aim of detecting those low energy Lorentz-violating effects. The electrodynamics sector of the SME 
has been studied in a broad class of experimental tests, as summarized in Ref.\ \cite{Russell}. For instance, constraints on the minimal electron sector
include those in Ref.\ \cite{electrons}, while ones in the minimal photon sector include those for the Chern-Simons term in Ref.\ \cite{photon3}
and those for propagation effects in Ref.\ \cite{photon4}.

At tree level, SME is well established. The one-loop renormalization of Extended QED (EQED) has already been performed and  the beta functions corresponding to 
each Lorentz-violating coupling \cite{Alan2} were calculated as well as for non-Abelian
extended theories in the pure Yang-Mills sector \cite{Colladay2}, in the electroweak \cite{Colladay3} and the strong \cite{Colladay4}
sectors. There are also several works regarding the radiative induction of Lorentz-violating terms \cite{CSI}. 

The chiral anomaly in the EQED has been studied recently \cite{Scarpelli} in the Fujikawa approach\cite{Fujikawa}. The anomaly cancellation conditions
are necessary even in the presence of Lorentz violation \cite{Colladay, Salvio}. 
In perturbative treatments, care must be exercised  in computing finite quantum corrections in extended such models. One reason is the choice of the treatment of infinities which is a subtle task in Lorentz-violating theories. Some of the Lorentz- and CPT-breaking terms contain dimensional specific objects 
like $\gamma_5$ matrices and Levi-Civita symbols. Consequently, conventional dimensional regularization \cite{DR, Bollini} is not suitable. Several frameworks have been proposed to overcome the difficulty in the treatment of $\gamma_5$ matrices in divergent amplitudes \cite{{Cynolter},{Tsai},{gamma5}}. 

Moreover it may occur {\it{a priori}}
undefined quantum corrections in Feynman diagram calculations, which
entail regularization dependence and spurious symmetry violation \cite{Jackiw}. This is particularly important in distinguishing between a genuine anomaly and an apparent violation of symmetry. Such arbitrariness being an artifact of divergences in perturbation theory should be fixed on symmetry grounds or physical constraints. When dimensional regularization and its extensions are ambiguous, it is convenient to employ a framework which, besides operating in the physical dimensional to avoid problems with dimensional specific theories, allows for a clear identification of arbitrary parameters. Therefore an useful tool is implicit regularization (IR) \cite{Nemes} in which such arbitrarinesses are expressed by finite differences between basic divergent loop integrals (independent of external momenta) having the same degree of divergence and different Lorentz structure which in turn can be expressed as surface terms (ST). The latter are usually multiplied by momentum routing dependent labels in the loops of a Feynman diagram. Some regularizations break momentum routing invariance (MRI),
a freedom authorized by energy-momentum conservation at the vertices of Feynman diagrams. 

The striking
connection between MRI and gauge symmetry was
realized long ago by t' Hooft and Veltman \cite{DR},  Jackiw \cite{MRI2} as well as
by Elias and collaborators \cite{MRI3}. In \cite{Adriano} it was established the interplay between the vanishing of ST and abelian gauge invariance to all orders in perturbation theory in the framework of IR: demanding MRI leads to preservation of Ward identities order by order in perturbation theory.
ST can be systematically derived at arbitrary loop order, complying with Lorentz invariance and unitarity as dictated by the local Bogoliubov's
R-operation based on the Bogoliubov-Parasiuk-Hepp-Zimmermann theorem \cite{{Cherchiglia},{BPHZ}}. Of course symmetry requirements may either be ensured by an
invariant regularization or imposed as constraint equations dictated by Ward-Slavnov-Taylor
identities order by order in the loops. Yet a little tedious from the calculational viewpoint, the latter procedure is perfectly sound for both anomaly free theories and models in which the quantum symmetry
breaking mechanism is well known. To cut a long story short, the choice of the regularization is a subtle task specially when we are studying anomalies and we want to find out whether 
a classical symmetry is also preserved at the quantum level. If an inadequate choice is made we can get spurious breaking terms which are not physical but are an artifact of the regularization itself.

In this work we evaluate the two- and three-point functions of the photon in the EQED completing the analysis of reference \cite{Alan2}, where the explicit 
computation of these diagrams was not considered due to finite radiative contributions which contain the dimensional specific gamma matrix $\gamma_5$ and need 
careful evaluation. We will explicitly show that MRI can be made manifest and the results will not be contaminated by arbitrary regularization dependent ST \cite
{Daniel}.  Instead,  we shall verify that Ward identities involving  two- and the three-point functions depend on such ST which, following \cite{Jackiw}, must be 
fixed either by phenomenology or symmetries of the underlying model. We conclude that MRI is a necessary and sufficient condition to establish gauge invariance 
in EQED. It has been shown that this invariance was also necessary to super-symmetry \cite{Adriano} and in the derivation of the Renormalization Group functions 
in scalar theories \cite{Cherchiglia}. Moreover when $\gamma_5$ matrices occur in Feynman diagram calculations care must be exercised even when we work in the 
physical dimension \cite{Cynolter}. A symmetrization of the traces involving such matrices is shown to be consistent with gauge invariance.

This work is organized as follows: in next section we present the EQED and its Feynman rules. In section
\ref{IReg} we present the calculational framework and notation we use to handle divergent amplitudes. In section \ref{sec4} we derive the gauge Ward identities 
and present their relation with MRI in section \ref{sec5}. We conclude in section \ref{conclusion} and leave a table of integrals for the appendix.

\section{The model and the amplitudes}

The EQED is a low-energy limit of the SME. It includes all Lorentz- and CPT-violating effects in some coefficients ($a, b, c,...$)
\cite{Alan2}:

\be
\mathcal{L} =\frac{1}{2}i \bar{\psi}\Gamma^{\mu}\overleftrightarrow{D}_{\mu}\psi- \bar{\psi}M\psi-\frac{1}{4}F^{\mu\nu}F_{\mu\nu}
-\frac{1}{4}(k_F)_{\kappa\lambda\mu\nu}F^{\mu\nu}F^{\kappa\lambda}+
\frac{1}{2}(k_{AF})^{\kappa}\epsilon_{\kappa\lambda\mu\nu}A^{\lambda}F^{\mu\nu}
\label{EQ1}
\ee

where $D_{\mu}\equiv \partial_{\mu}+ieA_{\mu}$ is the usual covariant derivative which couples the gauge field with matter,
\begin{align}
&\Gamma^{\nu}=\gamma^{\nu}+\Gamma^{\nu}_1,\nonumber\\
&\Gamma^{\nu}_1= c^{\mu\nu}\gamma_{\mu}+d^{\mu\nu}\gamma_5\gamma_{\mu}+e^{\nu}+i f^{\nu}\gamma_5+\frac{1}{2}g^{\lambda\mu\nu}
\sigma_{\lambda\mu}
\end{align}
and
\be
M= m+m_5 \gamma_5 +a^{\mu}\gamma_{\mu}+b_{\mu}\gamma_5\gamma^{\mu}+\frac{1}{2}H_{\mu\nu}\sigma^{\mu\nu}.
\ee

The coefficients $a_{\mu}$, $b_{\mu}$, $c_{\mu\nu}$, $d_{\mu\nu}$, $e_{\mu}$, $f_{\mu}$, $g_{\lambda\mu\nu}$, $H_{\mu\nu}$, $(k_F)_{\kappa\lambda\mu\nu}$ and 
$(k_{AF})_{\kappa}$ govern Lorentz violation and only the coefficients $a_{\mu}$, $b_{\mu}$, $e_{\mu}$, $f_{\mu}$, $g_{\lambda\mu\nu}$ and 
$(k_{AF})_{\kappa}$ govern CPT violation since the number of indices is odd.

\begin{figure}[!h]
 \includegraphics[trim=0mm 150mm 100mm 0mm, clip, scale=0.5]{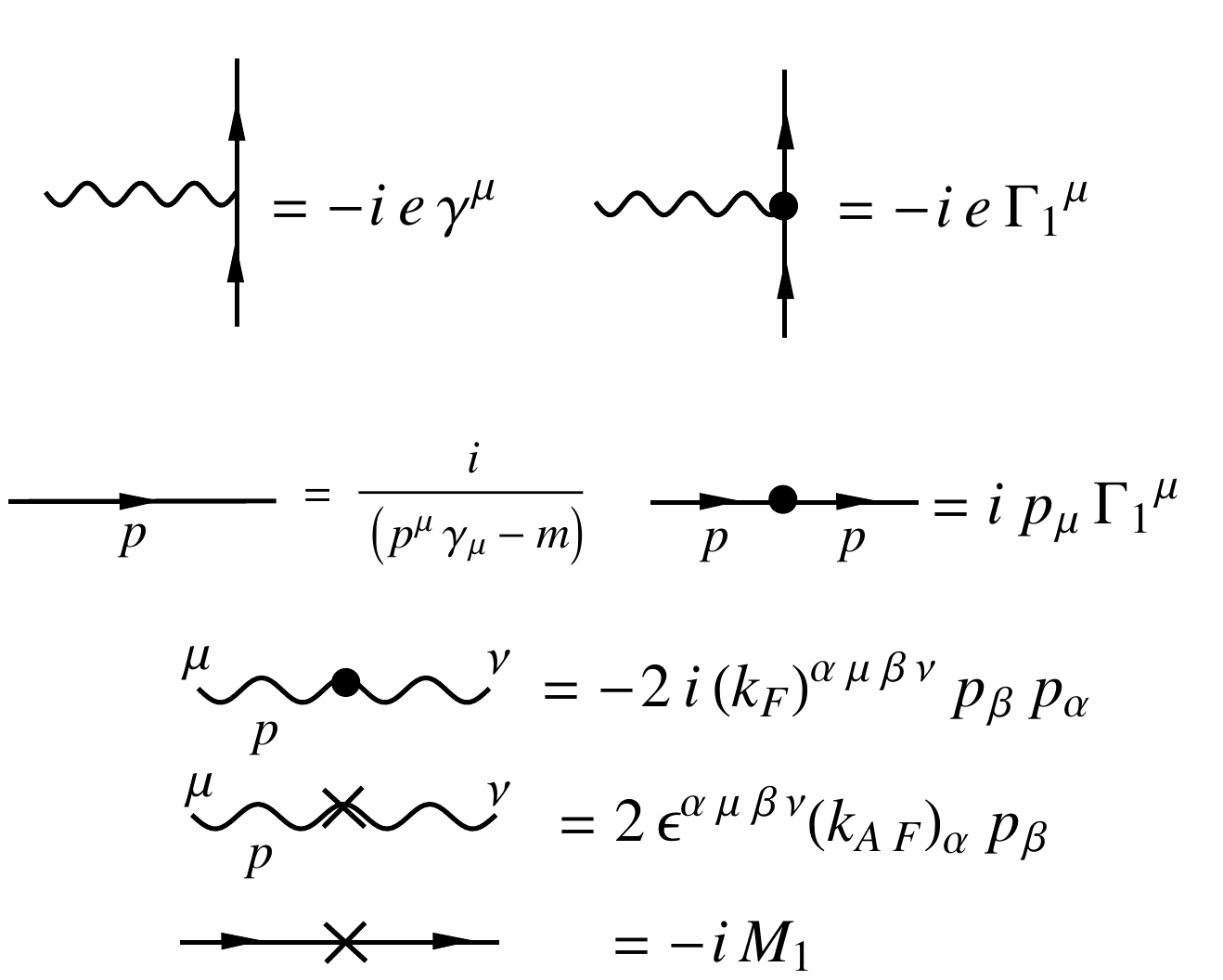}						
\caption{Feynman rules of the EQED.}     
\label{fig}
\end{figure}

We list the Feynman rules corresponding to theory (\ref{EQ1}) in figure \ref{fig}. We see in Lagrangian (\ref{EQ1}) that we have a general vertex with $\Gamma_\mu$
instead of just $\gamma_\mu$ and a general fermion propagator $\frac{i}{p_{\mu}\Gamma^\mu-M}$. However, considering the whole fermion propagator is a difficult 
task. Therefore, the dot and the cross insertions in figure \ref{fig} mean the leading order in Lorentz and CPT violation in the fermion propagator. 

Global gauge invariance of (\ref{EQ1}) leads to the classically conserved current  in this Lorentz- and CPT-violating version of QED:
\be
j^{\mu}=e\bar{\psi}\Gamma^{\mu}\psi.
\ee

\begin{figure}
 \includegraphics[trim=0mm 60mm 0mm 0mm,scale=0.8]{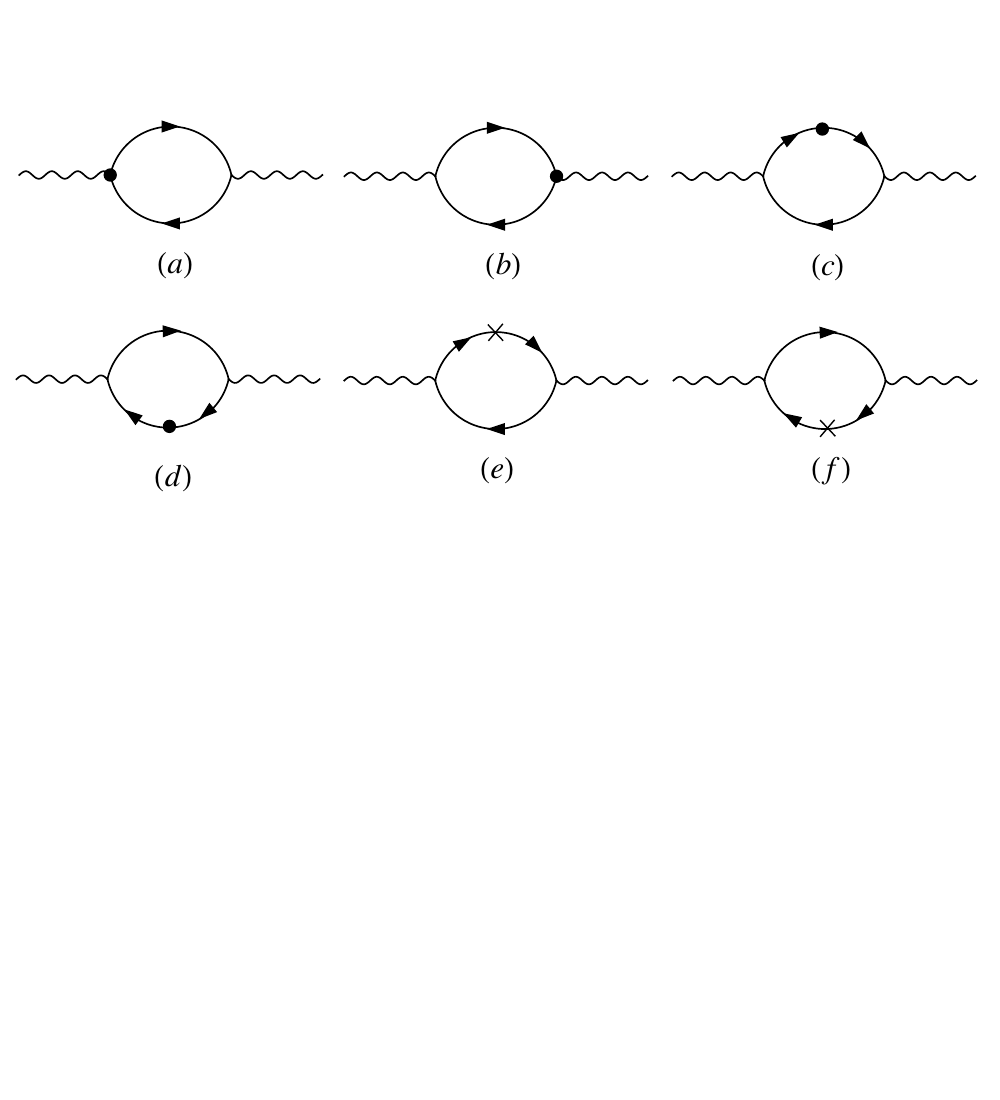}						
\caption{One-loop 2-point functions of the EQED.}     
\label{fig1}
\end{figure}

In order to verify whether or not the classical current $j^{\mu}$ is preserved at the quantum level, we have to obtain the gauge Ward identities coming from the evaluation of the diagrams of figure \ref{fig1} whose amplitudes are

\begin{figure}
 \includegraphics[scale=0.6]{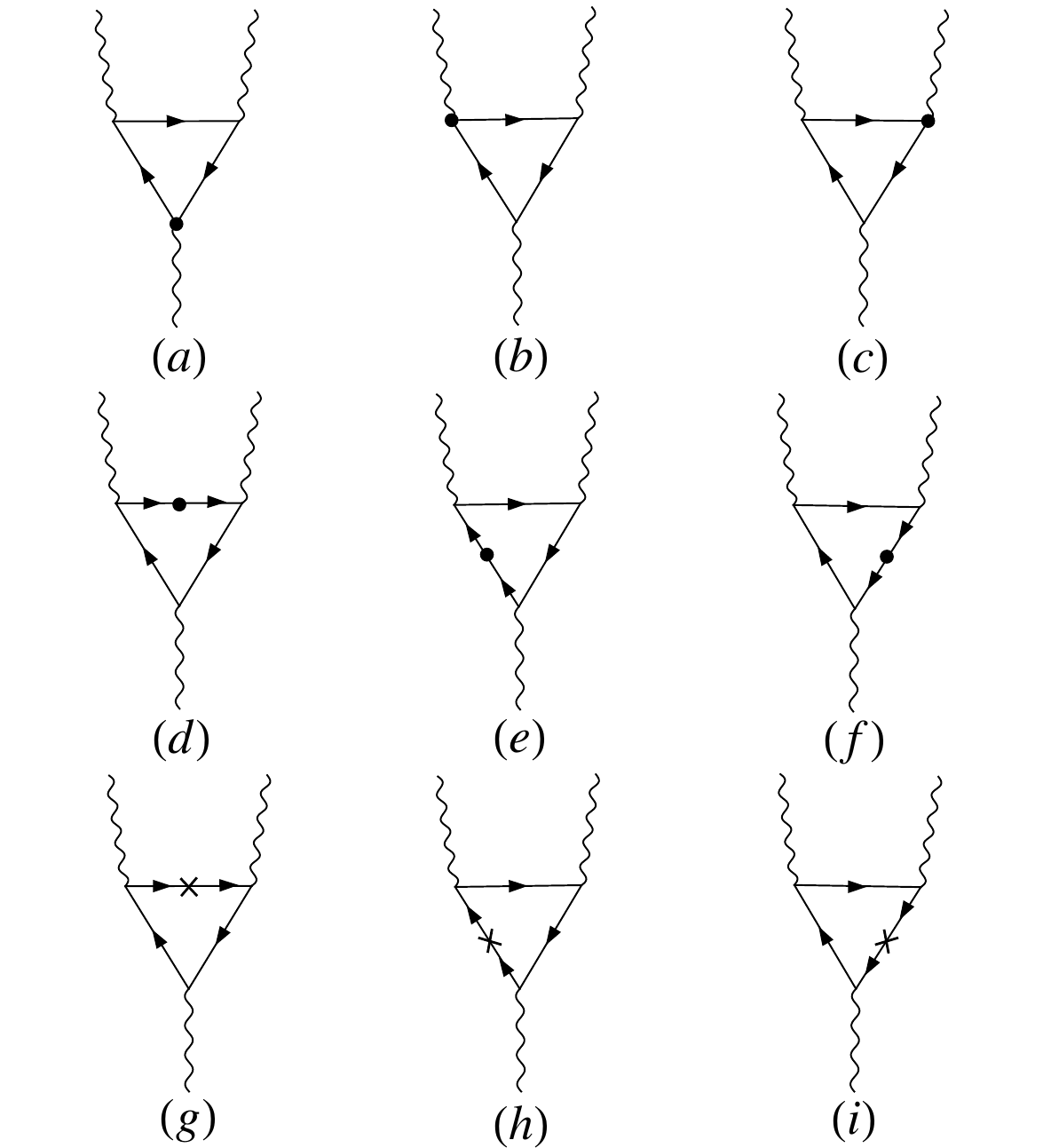}						
\caption{One-loop 3-point functions of the EQED.}     
\label{fig2}
\end{figure}

\be
\Pi^{\mu\nu}_{(a)}=-\int \frac{d^4 k}{(2\pi)^4} Tr \left[\Gamma_1^{\nu}\frac{1}{\slashed{k}-m}\gamma^{\mu}\frac{1}{\slashed{k}-\slashed
{p}-m}\right],
\label{(a)}
\ee

\be
\Pi^{\mu\nu}_{(b)}=-\int \frac{d^4 k}{(2\pi)^4} Tr \left[\gamma^{\nu}\frac{1}{\slashed{k}-m}\Gamma_1^{\mu}\frac{1}{\slashed{k}-\slashed{p}-m}\right],
\label{(b)}
\ee

\be
\Pi^{\mu\nu}_{(c)}=\int \frac{d^4 k}{(2\pi)^4} Tr \left[\gamma^{\nu}\frac{1}{\slashed{k}-m}\Gamma_1^{\lambda}k_{\lambda}\frac{1}{\slashed{k}-m}\gamma^{\mu}\frac{1}{\slashed{k}-\slashed{p}-m}\right],
\label{(c)}
\ee

\be
\Pi^{\mu\nu}_{(d)}=\int \frac{d^4 k}{(2\pi)^4} Tr \left[\gamma^{\nu}\frac{1}{\slashed{k}-m}\gamma^{\mu}\frac{1}{\slashed{k}-\slashed{p}-m}\Gamma_1^{\lambda}
(k_{\lambda}-p_{\lambda})\frac{1}{\slashed{k}-\slashed{p}-m}\right],
\label{(d)}
\ee

\be
\Pi^{\mu\nu}_{(e)}=-\int \frac{d^4 k}{(2\pi)^4} Tr \left[\gamma^{\nu} \frac{1}{\slashed{k}-m} M_1 \frac{1}{\slashed{k}-m} \gamma^{\mu} \frac{1}{\slashed{k}-
\slashed{p}-m} \right],
\label{(e)}
\ee

\be
\Pi^{\mu\nu}_{(f)}=-\int \frac{d^4 k}{(2\pi)^4} Tr \left[\gamma^{\nu} \frac{1}{\slashed{k}-m} \gamma^{\mu} \frac{1}{\slashed{k}-
\slashed{p}-m} M_1 \frac{1}{\slashed{k}-\slashed{p}-m} \right].
\label{(f)}
\ee

There is also a Ward identity coming from the 3-point functions of figure \ref{fig2}. These amplitudes read

\be
T^{\mu\nu\alpha}_{(a)}=-\int \frac{d^4 k}{(2\pi)^4} Tr \left[\gamma^{\mu}\frac{1}{\slashed{k}-m}\gamma^{\nu}\frac{1}{\slashed{k}+\slashed
{q}-m}\Gamma_1^{\alpha}\frac{1}{\slashed{k}-\slashed{p}-m}\right],
\label{A1}
\ee

\be
T^{\mu\nu\alpha}_{(b)}=-\int \frac{d^4 k}{(2\pi)^4} Tr \left[\gamma^{\mu}\frac{1}{\slashed{k}-m}\Gamma_1^{\nu}\frac{1}{\slashed{k}+\slashed
{q}-m}\gamma^{\alpha}\frac{1}{\slashed{k}-\slashed{p}-m}\right],
\label{A2}
\ee

\be
T^{\mu\nu\alpha}_{(c)}=-\int \frac{d^4 k}{(2\pi)^4} Tr \left[\Gamma_1^{\mu}\frac{1}{\slashed{k}-m}\gamma^{\nu}\frac{1}{\slashed{k}+\slashed
{q}-m}\gamma^{\alpha}\frac{1}{\slashed{k}-\slashed{p}-m}\right],
\label{A3}
\ee

\be
T^{\mu\nu\alpha}_{(d)}=\int \frac{d^4 k}{(2\pi)^4} Tr \left[\gamma^{\mu}\frac{1}{\slashed{k}-m}\Gamma_1^{\lambda}k_{\lambda}\frac{1}{
\slashed{k}-m}\gamma^{\nu}\frac{1}{\slashed{k}+\slashed{q}-m}\gamma^{\alpha}\frac{1}{\slashed{k}-\slashed{p}-m}\right],
\label{A4}
\ee

\be
T^{\mu\nu\alpha}_{(e)}=\int \frac{d^4 k}{(2\pi)^4} Tr \left[\gamma^{\mu}\frac{1}{\slashed{k}-m}\gamma^{\nu}\frac{1}{\slashed{k}+\slashed
{q}-m}\Gamma_1^{\lambda}(k+q)_{\lambda}\frac{1}{\slashed{k}+\slashed{q}-m}\gamma^{\alpha}\frac{1}{\slashed{k}-\slashed{p}-m}\right],
\label{A5}
\ee

\be
T^{\mu\nu\alpha}_{(f)}=\int \frac{d^4 k}{(2\pi)^4} Tr \left[\gamma^{\mu}\frac{1}{\slashed{k}-m}\gamma^{\nu}\frac{1}{\slashed{k}+\slashed
{q}-m}\gamma^{\alpha}\frac{1}{\slashed{k}-\slashed{p}-m}\Gamma_1^{\lambda}(k-p)_{\lambda}\frac{1}{\slashed{k}-\slashed{p}-m}\right].
\label{A6}
\ee

\be
T^{\mu\nu\alpha}_{(g)}=-\int \frac{d^4 k}{(2\pi)^4} Tr \left[\gamma^{\mu}\frac{1}{\slashed{k}-m}M_1\frac{1}{\slashed{k}-m}\gamma^{\nu}\frac{1}{\slashed{k}+\slashed{q}-m}\gamma^{\alpha}\frac{1}{\slashed{k}-\slashed{p}-m}\right],
\label{A7}
\ee

\be
T^{\mu\nu\alpha}_{(h)}=-\int \frac{d^4 k}{(2\pi)^4} Tr \left[\gamma^{\mu}\frac{1}{\slashed{k}-m}\gamma^{\nu}\frac{1}{\slashed{k}+\slashed
{q}-m}M_1\frac{1}{\slashed{k}+\slashed{q}-m}\gamma^{\alpha}\frac{1}{\slashed{k}-\slashed{p}-m}\right],
\label{A8}
\ee

\be
T^{\mu\nu\alpha}_{(i)}=-\int \frac{d^4 k}{(2\pi)^4} Tr \left[\gamma^{\mu}\frac{1}{\slashed{k}-m}\gamma^{\nu}\frac{1}{\slashed{k}+\slashed
{q}-m}\gamma^{\alpha}\frac{1}{\slashed{k}-\slashed{p}-m}M_1\frac{1}{\slashed{k}-\slashed{p}-m}\right].
\label{A9}
\ee

Gauge invariance holds at one-loop order if the following vectorial Ward identities hold:

\begin{eqnarray}
&p_{\mu}(\Pi^{\mu\nu}_{(a)}+\Pi^{\mu\nu}_{(b)}+\Pi^{\mu\nu}_{(c)}+\Pi^{\mu\nu}_{(d)}+\Pi^{\mu\nu}_{(e)}+
\Pi^{\mu\nu}_{(f)}) =0 \\ 
\label{GWI0} 
&(p+q)_{\alpha}(T^{\mu\nu\alpha}_{(a)}+T^{\mu\nu\alpha}_{(b)}+T^{\mu\nu\alpha}_{(c)}+T^{\mu\nu\alpha}_{(d)}+T^{\mu\nu\alpha}_{(e)}
+T^{\mu\nu\alpha}_{(f)}+T^{\mu\nu\alpha}_{(g)}+T^{\mu\nu\alpha}_{(h)}+T^{\mu\nu\alpha}_{(i)})=0
\label{GWI} 
\end{eqnarray}

There are also the corresponding crossed diagrams of figure \ref{fig2} to be considered. Taking for instance, diagram $T^{\mu\nu\alpha}_{(a)}$ with its crossed 
diagram, we have

\begin{align}
&T^{\mu\nu\alpha}_{(a)}(p,q)+T^{\nu\mu\alpha}_{(a)}(q,p)=\nonumber\\
=&-\int \frac{d^4 k}{(2\pi)^4} Tr \left[\gamma^{\mu}\frac{1}{\slashed{k}-m}\gamma^{
\nu}\frac{1}{\slashed{k}+\slashed{q}-m}\Gamma_1^{\alpha}\frac{1}{\slashed{k}-\slashed{p}-m}\right]+\int \frac{d^4 k}{(2
\pi)^4} Tr \left[\frac{1}{\slashed{k}-\slashed{p}+m}\Gamma_1^{\alpha}\frac{1}{\slashed{k}+\slashed{q}+m}\gamma^{\nu}\frac{1}{\slashed
{k}+m}\gamma^{\mu}\right].
\end{align}

We can take the transpose of the square bracket of second term and use that $(\gamma^{\mu})^T=-C^{-1}\gamma^{\mu}C$ to obtain the following term in the numerator:
\be
Tr[\gamma^{\mu}(\slashed{k}+m)\gamma^{\nu}(\slashed{k}+\slashed{q}+m)(-C(\Gamma_1^{\alpha})^T C^{-1}-\Gamma_1^{\alpha})
(\slashed{k}-\slashed{p}+m)],
\ee
where
\be
-C(\Gamma_1^{\nu})^T C^{-1}=c^{\mu\nu}\gamma_{\mu}-d^{\mu\nu}\gamma_5\gamma_{\mu}-e^{\nu}-i f^{\nu}
\gamma_5+\frac{1}{2}g^{\lambda\mu\nu}\sigma_{\lambda\mu}.
\ee

Therefore, the overall effect of the crossed diagrams  is to cancel the C-preserving part of $\Gamma_1^{\mu}$ as is presented in \cite{Alan2}. Then we may simply replace $\Gamma_1^{\alpha}$ with $\bar{\Gamma}_1^{\alpha}=C(\Gamma_1^{\alpha})^T C^{-1}+\Gamma_1^{\alpha}$ in amplitudes (\ref{A1})-(\ref{A6}) to consider the crossed diagrams. 


Of course, the same idea can be applied to the the insertion $M_1$ in diagrams $(g)$, $(h)$ and $(i)$. However, the sum of one diagram with its crossed diagram 
would lead to $M_1-C(M_1)^T C^{-1}\equiv \bar{M}_1$ in the numerator. Consequently, $m_5$ and $b_{\mu}$ coefficients are the ones which vanish.

\section{Basic Divergent Integrals and Surface Terms}
\label{IReg}

Here we briefly describe the  framework  \cite{Nemes} to handle the divergent integrals in four dimensions which appear in the amplitudes of the previous section and establish some notation.  In this scheme, we assume that integrals are regularized by an 
implicit regulator $\Lambda$ just to justify algebraic operations within the integrands. We then use, for instance, the following identity 
\be
\int_k\frac{1}{(k+p)^2-m^2}=\int_k\frac{1}{k^2-m^2}-\int_k\frac{(p^2+2p\cdot k)}{(k^2-m^2)[(k+p)^2-m^2]},
\label{2.1}
\ee
where $\int_k\equiv\int^\Lambda\frac{d^4 k}{(2\pi)^4}$ in order to separate basic divergent integrals from the finite part. They are defined as follows:
\begin{equation}
I^{\mu_1 \cdots \mu_{2n}}_{log}(m^2)\equiv \int_k \frac{k^{\mu_1}\cdots k^{\mu_{2n}}}{(k^2-m^2)^{2+n}}
\end{equation}
and
\begin{equation}
I^{\mu_1 \cdots \mu_{2n}}_{quad}(m^2)\equiv \int_k \frac{k^{\mu_1}\cdots k^{\mu_{2n}}}{(k^2-m^2)^{1+n}}.
\end{equation}

The basic divergences with Lorentz indices can be judiciously combined as differences between integrals with the same superficial degree 
of divergence, according to the equations below, which define the ST \footnote{The Lorentz indices between brackets stand for permutations, {\it i.e.}, 
$A^{\{\alpha_1\cdots\alpha_n}B^{\beta_1\cdots\beta_n\}}=A^{\alpha_1\cdots\alpha_{n}}B^{\beta_1\cdots\beta_n}$ 
+ sum over permutations between the two sets of indices $\alpha_1\cdots\alpha_{n}$ and $\beta_1\cdots\beta_n$.}:
\begin{eqnarray}
&\Upsilon^{\mu \nu}_{2w}=  g^{\mu \nu}I_{2w}(m^2)-2(2-w)I^{\mu \nu}_{2w}(m^2)=\upsilon_{2w}g^{\mu \nu},
\label{dif1}\\
\nonumber\\
&\Xi^{\mu \nu \alpha \beta}_{2w}=  g^{\{ \mu \nu} g^{ \alpha \beta \}}I_{2w}(m^2)
 -4(3-w)(2-w)I^{\mu \nu \alpha \beta }_{2w}(m^2)=  \xi_{2w}g^{\{{\mu \nu}} g^{{\alpha \beta}\}},
\label{diff2}\\
\nonumber\\
&\Sigma^{\mu \nu \alpha \beta \gamma \delta}_{2w} = g^{\{\mu \nu} g^{ \alpha \beta} g^ {\gamma \delta \}}I_{2w}(m^2)
-8(4-w)(3-w)(2-w) I^{\mu \nu \alpha \beta \gamma \delta}_{2w}(m^2)= \sigma_{2w} g^{\{\mu \nu} g^{ \alpha \beta} g^ {\gamma \delta \}}.
\label{dif3}
\end{eqnarray} 

In the expressions above, $2w$ is the degree of divergence of the integrals and  for the sake of brevity, we substitute the subscripts 
$log$ and $quad$ by $0$ and $2$, respectively. Surface terms can be conveniently written as integrals of total 
derivatives, namely
\begin{eqnarray}
\upsilon_{2w}g^{\mu \nu}= \int_k\frac{\partial}{\partial k_{\nu}}\frac{k^{\mu}}{(k^2-m^2)^{2-w}}, \nonumber \\
\label{ts1}
\end{eqnarray}
\begin{eqnarray}
(\xi_{2w}-v_{2w})g^{\{{\mu \nu}} g^{{\alpha \beta}\}}= \int_k\frac{\partial}{\partial 
k_{\nu}}\frac{2(2-w)k^{\mu} k^{\alpha} k^{
\beta}}{(k^2-m^2)^{3-w}}.
\label{ts2}
\end{eqnarray}
and
\begin{eqnarray}
(\sigma_{2w}-\xi_{2w})g^{\{ \mu \nu} g^{ \alpha \beta} g^ {\gamma \delta \}} =
\int_k\frac{\partial}{\partial k_{\nu}}\frac{4(3-w)(2-w)k^{\mu} k^{\alpha} k^{\beta} k^{\gamma} k^{\delta}}{(k^2-m^2)^{4-w}}.
\label{ts3}
\end{eqnarray}

Equations (\ref{dif1})-(\ref{dif3}) are in principle arbitrary and regularization dependent. They can be shown to vanish in usual dimensional regularization. We 
leave these terms unevaluated until the end of the calculation to be fixed  on symmetry grounds or phenomenology, when it is the case \cite{Jackiw}.
When a quantum breaking runs over different Ward identities \cite{{Adriano},{LEONARDO}}, {\it e.g.}, the AVV triangle anomaly, they play a fundamental role in a 
democratic display of the anomaly.

\section{Evaluation of the diagrams}
\label{sec4}

\subsection{Two-point function}

We start this section evaluating the Ward identity coming from the two-point function presented in figure \ref{fig1}. The left hand side of equation (\ref{GWI0}) reads:

\begin{align}
&p_{\mu}\sum^{f}_{i=a}\Pi^{\mu\nu}_{(i)}(p)= \Big[- \int_k Tr \left(\Gamma_1^{\nu}\frac{1}{\slashed{k}-\slashed{p}-m}\right)+\int_k Tr \left(\Gamma_1^{\nu}\frac
{1}{\slashed{k}-m}\right)\Big]-\Big[\int_k Tr \left(\gamma^{\nu}\frac{1}{\slashed{k}-m}p_{\mu}\Gamma_1^{\mu}\frac{1}{\slashed{k}-\slashed{p}-m}
\right)\Big]+\nonumber\\&+\Big[\int_k Tr \left(\gamma^{\nu}\frac{1}{\slashed{k}-m}k_{\lambda}\Gamma_1^{\lambda}\frac{1}{\slashed{k}-\slashed{p}-m}\right)-\int_k 
Tr \left(\gamma^{\nu}\frac{1}{\slashed{k}-m}k_{\lambda}\Gamma_1^{\lambda}\frac{1}{\slashed{k}-m}\right)\Big]+\nonumber\\&+\Big[\int_k Tr \left(\gamma^{\nu}\frac
{1}{\slashed{k}-\slashed{p}-m}(k-p)_{\lambda}\Gamma_1^{\lambda}\frac{1}{\slashed{k}-\slashed{p}-m}\right)-\int_k Tr \left(\gamma^{\nu}\frac{1}{\slashed
{k}-m}(k-p)_{\lambda}\Gamma_1^{\lambda}\frac{1}{\slashed{k}-\slashed{p}-m}\right)\Big]+\nonumber\\&+\Big[-\int_k Tr \left(\gamma^{\nu}\frac{1}{\slashed{k}-m}M_1
\frac{1}{\slashed{k}-\slashed{p}-m}\right)+\int_k Tr \left(\gamma^{\nu}\frac{1}{\slashed{k}-m}M_1\frac{1}{
\slashed{k}-m}\right)\Big]+\nonumber\\&+\Big[-\int_k Tr \left(\gamma^{\nu}\frac{1}{\slashed{k}-\slashed{p}-m}M_1\frac{1}{\slashed{k}-
\slashed{p}-m}\right)+\int_k Tr \left(\gamma^{\nu}\frac{1}{\slashed{k}-m}M_1\frac{1}{\slashed{k}-\slashed{p}-m}\right)\Big]
\label{WI1}
\end{align}

Diagrams in equations (\ref{(a)})-(\ref{(f)}) correspond to each square bracket of equation (\ref{WI1}), respectively. We have used $\slashed{p}=\slashed{k}-m-(\slashed{k}-\slashed{p}-m)$ in order to exclude one propagator and 
separate in two parts some diagrams. The summation of the first term of diagram $(c)$, the second term of diagram $(d)$ and  
diagram $(b)$ cancels. Likewise, the first term of diagram $(e)$ cancels the second term of diagram $(f)$. After trace algebra, the integrals are evaluated as displayed in the appendix. Putting the results together, we verify that it is transverse up to surface terms:

\begin{align}
&p_{\mu}\sum^{f}_{i=a}\Pi^{\mu\nu}_{(i)}(p)= 4(12 c^{\alpha\beta}p_{\alpha}p_{\beta}p^{\nu}+4 p^{2}c^{\nu\alpha}p_{\alpha}+4 p^{2}c^{\alpha\nu}p_{\alpha})(\xi_0-\upsilon_0)+4(p^2 (m e^{\nu}-a^{\nu})+2(m e\cdot p-a\cdot p) p^{\nu})\times \nonumber\\&\times (2\upsilon_0-\xi_0)+4(c^{\nu\alpha}p_{\alpha
}+c^{\alpha\nu}p_{\alpha})(2\upsilon_2-\xi_2-p^2\sigma_0)-8 c^{\alpha\beta}p_{\alpha}p_{\beta}p^{\nu}\sigma_0
\label{WIPI}
\end{align}

The result above depends on the arbitrary logarithmic surface terms $\upsilon_0$, $\xi_0$ and $\sigma_0$ and 
on the arbitrary quadratic surface terms $\upsilon_2$ and $\xi_2$. As we discussed earlier  these
terms expressed by differences between two divergent quantities are regularization dependent and must be 
fixed on symmetry grounds \cite{Jackiw}. For instance, gauge invariance fixes those arbitrary terms (see \cite{{Adriano},{Cabral}}).
However, since we want to prove whether gauge symmetry holds in EQED beyond tree level, is not possible to require this symmetry to fix the arbitrarinesses since
we do not know if it holds {\it a priori}. Thus, we have to require another constraint to fix these arbitrary terms.
We require Momentum Routing Invariance (MRI) of the diagrams. That means physical results should not depend on the way we label the internal momenta as long as 
we respect the energy-momentum conservation at each vertex. We can take advantage of this feature when we do not have another symmetry to require in order to
fix arbitrarinesses. In theories with only scalar fields we have the lack of gauge symmetry and MRI is the only condition we have to fix surface terms. We will 
show that MRI also fixes such surface terms {\it{leading to}} gauge invariance in EQED. The proof 
that MRI and abelian gauge invariance are interconnected has been established for QED in \cite{Adriano}. We will discuss all this further in section \ref{sec5} 
after evaluating the three-point functions. 

It is noteworthy that we have the combination $m e^{\nu}-a^{\nu}$ in eq. (\ref{WIPI}) that is not expected to be observable since it can be eliminated by
a field redefinition \cite{Colladay}. However, we also have the symmetric part of $c_{\mu\nu}$ coefficient that is in principle observable. A spurious value for 
the surface terms would lead to the wrong conclusion of a gauge anomaly expressed by either observable or non-observable coefficients.

\subsection{Three-point function}

We proceed to evaluate the left hand side of equation (\ref{GWI}):

\begin{align}
&(p+q)_{\alpha}\sum^{i}_{j=a}T^{\mu\nu\alpha}_{(j)}(p,q)= \Big[- \int_k Tr \left(\gamma^{\mu}\frac{1}{\slashed{k}-m}\gamma^{\nu
}\frac{1}{\slashed{k}+\slashed{q}-m}\bar{\Gamma}_1^{\alpha}(p+q)_{\alpha}\frac{1}{\slashed{k}-\slashed{p}-m}\right)\Big]+\Big[-\int_k Tr 
\left(\gamma^{\mu}\frac{1}{\slashed{k}-m}\bar{\Gamma}_1^{\nu}\frac{1}{\slashed{k}-\slashed{p}-m}\right)+\nonumber\\&+\int_k Tr \left(\gamma^{
\mu}\frac{1}{\slashed{k}-m}\bar{\Gamma}_1^{\nu}\frac{1}{\slashed{k}+\slashed{q}-m} \right) \Big]+\Big[-\int_k Tr 
\left(\bar{\Gamma}_1^{\mu}\frac{1}{\slashed{k}-m}\gamma^{\nu}\frac{1}{\slashed{k}-\slashed{p}-m}\right)+\int_k Tr \left(\bar{\Gamma}_1^{
\mu}\frac{1}{\slashed{k}-m}\gamma^{\nu}\frac{1}{\slashed{k}+\slashed{q}-m} \right) \Big]+\nonumber\\&+
\Big[\int_k Tr \left(\gamma^{\mu}\frac{1}{\slashed{k}-m}\bar{\Gamma}_1^{\lambda}k_{\lambda}\frac{1}{\slashed{k}-m}\gamma^{\nu}\frac{1}{\slashed
{k}-\slashed{p}-m}\right)-\int_k Tr \left(\gamma^{\mu}\frac{1}{\slashed{k}-m}\bar{\Gamma}_1^{\lambda}k_{\lambda}\frac{1}{\slashed{k}-m}\gamma^{
\nu}\frac{1}{\slashed{k}+\slashed{q}-m}\right)\Big]+\nonumber\\&+\Big[\int_k Tr \left(\gamma^{\mu}\frac{1}{\slashed{k}-m}\gamma^{\nu
}\frac{1}{\slashed{k}+\slashed{q}-m}\bar{\Gamma}_1^{\lambda}(k+q)_{\lambda}\frac{1}{\slashed{k}-\slashed{p}-m}\right)-\int_k Tr \left(\gamma_{
\mu}\frac{1}{\slashed{k}-m}\gamma^{\nu}\frac{1}{\slashed{k}+\slashed{q}-m}\bar{\Gamma}_1^{\lambda}(k+q)_{\lambda}\frac{1}{\slashed{k}+\slashed
{q}-m}\right)\Big]+\nonumber\\&+\Big[-\int_k Tr \left(\gamma^{\mu}\frac{1}{\slashed{k}-m}\gamma^{\nu
}\frac{1}{\slashed{k}+\slashed{q}-m}\bar{\Gamma}_1^{\lambda}(k-p)_{\lambda}\frac{1}{\slashed{k}-\slashed{p}-m}\right)+\int_k Tr \left(\gamma_{
\mu}\frac{1}{\slashed{k}-m}\gamma^{\nu}\frac{1}{\slashed{k}-\slashed{p}-m}\bar{\Gamma}_1^{\lambda}(k-p)_{\lambda}\frac{1}{\slashed{k}-\slashed
{p}-m}\right) \Big]
+\nonumber\\&+
\Big[-\int_k Tr \left(\gamma^{\mu}\frac{1}{\slashed{k}-m}\bar{M}_1\frac{1}{\slashed{k}-m}\gamma^{\nu}\frac{1}{\slashed
{k}-\slashed{p}-m}\right)+\int_k Tr \left(\gamma^{\mu}\frac{1}{\slashed{k}-m}\bar{M}_1\frac{1}{\slashed{k}-m}\gamma^{
\nu}\frac{1}{\slashed{k}+\slashed{q}-m}\right)\Big]+\nonumber\\&+\Big[-\int_k Tr \left(\gamma^{\mu}\frac{1}{\slashed{k}-m}\gamma^{\nu
}\frac{1}{\slashed{k}+\slashed{q}-m}\bar{M}_1\frac{1}{\slashed{k}-\slashed{p}-m}\right)+\int_k Tr \left(\gamma_{
\mu}\frac{1}{\slashed{k}-m}\gamma^{\nu}\frac{1}{\slashed{k}+\slashed{q}-m}\bar{M}_1\frac{1}{\slashed{k}+\slashed
{q}-m}\right)\Big]+\nonumber\\&+\Big[\int_k Tr \left(\gamma^{\mu}\frac{1}{\slashed{k}-m}\gamma^{\nu
}\frac{1}{\slashed{k}+\slashed{q}-m}\bar{M}_1\frac{1}{\slashed{k}-\slashed{p}-m}\right)-\int_k Tr \left(\gamma_{
\mu}\frac{1}{\slashed{k}-m}\gamma^{\nu}\frac{1}{\slashed{k}-\slashed{p}-m}\bar{M}_1\frac{1}{\slashed{k}-\slashed
{p}-m}\right) \Big].
\label{soma}
\end{align}

Diagrams represented by equations (\ref{A1})-(\ref{A9}) correspond to each square bracket of equation (\ref{soma}), respectively. Again we use
algebra like $\slashed{p}+\slashed{q}=\slashed{k}+\slashed{q}-m-(\slashed{k}-\slashed{p}-m)$ in order to exclude one propagator and split some amplitudes into two parts. The summation of the first term of diagram $(e)$, the first term of diagram $(f)$ and  
diagram $(a)$ cancels. The first term of diagram $(h)$ cancels the first term of diagram $(i)$. For the remaining (divergent) terms, we do not perform  shifts of integration variable as they lead to extra surface terms: the final result is finite, it is important not to evaluate arbitrary undetermined terms especially those involving quadratically divergent surface terms.  

Moreover one has to be careful in evaluating  $Tr[\gamma ^{\mu }\gamma ^{\beta }\gamma ^{\nu }\gamma ^{\xi 
}\gamma ^{\alpha}\gamma ^{\lambda }\gamma ^5]$ that we find in these divergent amplitudes. It is possible to use the following identity to reduce the number of Dirac
$\gamma$ matrices:
\be
\gamma ^{\mu }\gamma ^{\beta }\gamma ^{\nu }= g^{\mu \beta} \gamma^{\nu}+g^{\nu \beta} \gamma^{\mu}-g^{\mu \nu} \gamma^{\beta}-i 
\epsilon^{\mu \beta \nu \rho}\gamma_{\rho}\gamma_5.
\label{gamma}
\ee

Using equation (\ref{gamma}), $Tr[\gamma ^{\mu }\gamma ^{\beta }\gamma ^{\nu }\gamma ^{\xi }\gamma ^5]=4i\epsilon^{\mu \beta \nu \xi}$ 
and $\gamma_5\gamma^{\rho}\gamma_5=-\gamma^{\rho}$ we end up with the following result

\be
Tr[\gamma ^{\mu }\gamma ^{\beta }\gamma ^{\nu }\gamma ^{\xi }\gamma ^{\alpha}\gamma ^{\lambda }\gamma ^5]=4i(g^{\beta \mu}\epsilon^{\nu 
\xi \alpha \lambda}+g^{\beta \nu}\epsilon^{\mu \xi \alpha \lambda}-g^{\mu \nu}\epsilon^{\beta \xi \alpha \lambda}
-g^{\lambda \alpha}\epsilon^{\mu \beta \nu \xi}+g^{\xi \lambda}\epsilon^{\mu \beta \nu \alpha}-g^{\xi \alpha}\epsilon^{\mu \beta \nu 
\lambda}).
\label{trace}
\ee

However, it is completely arbitrary which three $\gamma$ matrices we pick up to apply equation (\ref{gamma}). A different choice would 
give the result of equation (\ref{trace}) with Lorentz indices permuted. Furthermore, equation $\{\gamma_5,\gamma_{\mu}\}=0$ should be 
avoided inside a divergent integral since this operation seems to fix a value for the surface term which explains
why it breaks gauge invariance in some calculations \cite{Tsai}. This approach was also discussed in \cite{Cynolter}. Therefore, we have to consider all possible permutations of Lorentz indices in equation (\ref
{trace}) to yield  \cite{{Cynolter},{Perez}}:
\begin{align}
&\frac{-i}{4}Tr[\gamma ^{\mu }\gamma ^{\nu }\gamma ^{\alpha }\gamma ^{\beta }\gamma ^{\gamma }\gamma ^{\delta }\gamma ^5]=-g^{\alpha 
\beta } \epsilon ^{\gamma \delta \mu \nu }+g^{\alpha \gamma } \epsilon ^{\beta \delta \mu \nu }-g^{\alpha \delta } \epsilon 
^{\beta \gamma \mu \nu }-g^{\alpha \mu } \epsilon ^{\beta \gamma \delta \nu }+g^{\alpha \nu } \epsilon ^{\beta \gamma \delta \mu }-g^{
\beta \gamma } \epsilon ^{\alpha \delta \mu \nu }+g^{\beta \delta } \epsilon ^{\alpha \gamma \mu \nu }+\nonumber\\&+ g^{\beta \mu } 
\epsilon ^{\alpha \gamma \delta \nu }-g^{\beta \nu } \epsilon ^{\alpha \gamma \delta \mu }-g^{\gamma \delta } \epsilon ^{\alpha \beta \mu
\nu }-g^{\gamma \mu } \epsilon ^{\alpha \beta \delta \nu }+g^{\gamma \nu } \epsilon ^{\alpha \beta \delta \mu }+g^{\delta 
\mu } \epsilon ^{\alpha \beta \gamma \nu }-g^{\delta \nu } \epsilon ^{\alpha \beta \gamma \mu }-g^{\mu \nu } \epsilon ^{\alpha \beta 
\gamma \delta }.
\label{traco2}
\end{align}
After taking the remaining traces of equation (\ref{soma}), we regularize the divergent integrals using the method of section \ref
{IReg}. We list all the regularized integrals in the appendix. The result of the Ward identity is

\begin{align}
&(p+q)_{\alpha}\sum^{i}_{j=a}T^{\mu\nu\alpha}_{(j)}= 4i(d_{\lambda\gamma}+d_{\gamma\lambda})p^{\lambda}p_{\alpha}
\epsilon^{\alpha\gamma\mu\nu}(\upsilon_0-\xi_0)+ 8m(p^{\nu}e^{\mu}+p^{\mu}e^{\nu}+g^{\mu\nu}e\cdot p)(2
\upsilon_0-\xi_0)+\nonumber\\&+ 8(p^{\nu}a^{\mu}+p^{\mu}a^{\nu}+g^{\mu\nu}a\cdot p)(2\upsilon_0-\xi_0)-(-p \rightarrow q).
\end{align}

As in the previous subsection, the Ward identity depends on the arbitrary surface terms, only the logarithmic ones this time since the amplitudes are linearly 
divergent. Therefore we notice, once again, that gauge invariance is broken by arbitrary terms, which should be fixed by demanding MRI as we demonstrate in the next section.

\section{The gauge and momentum routing invariance relation}
\label{sec5}

As discussed in the previous section, Ward identities for two and three-point functions in EQED are only broken by regularization-dependent objects. This feature  allows us to draw already some conclusions:
first, that \textbf{no} anomaly occurs in EQED (at least at one-loop order). The reason lies on the appearance of only ST, which are regularization-dependent. In other words, they only parametrize breakings by regularization methods themselves (for example, the cut-off regularization always evaluates surface terms to finite values, explaining why this formalism is known to break gauge invariance). If an anomaly occurs, the Ward identity should evaluate to \textbf{finite} terms (which would represent the quantum breaking) plus surface terms. One could now wonder if, by imposing some other constraint, the surface terms could be fixed. The answer is positive and that constraint is MRI as we show bellow.

MRI is a feature that should be shared by all theories since it means physical results should not depend on the way we label the internal momenta of Feynman 
diagrams as long as we respect energy-momentum conservation at each vertex. The non-trivial fact to be emphasized is that, as showed in \cite{Adriano}, MRI 
breakings \textbf{always} manifest themselves by the appearance of surface terms. Since gauge invariance is also broken by the same objects, one can speculate 
that it is related to MRI. This statement was showed to be true for Standard QED in \cite{Adriano}. Here we will sketch how the same proof can be applied for EQED. 

\begin{figure}[!h]
 \includegraphics[trim=0mm 90mm 0mm 100mm,scale=0.3]{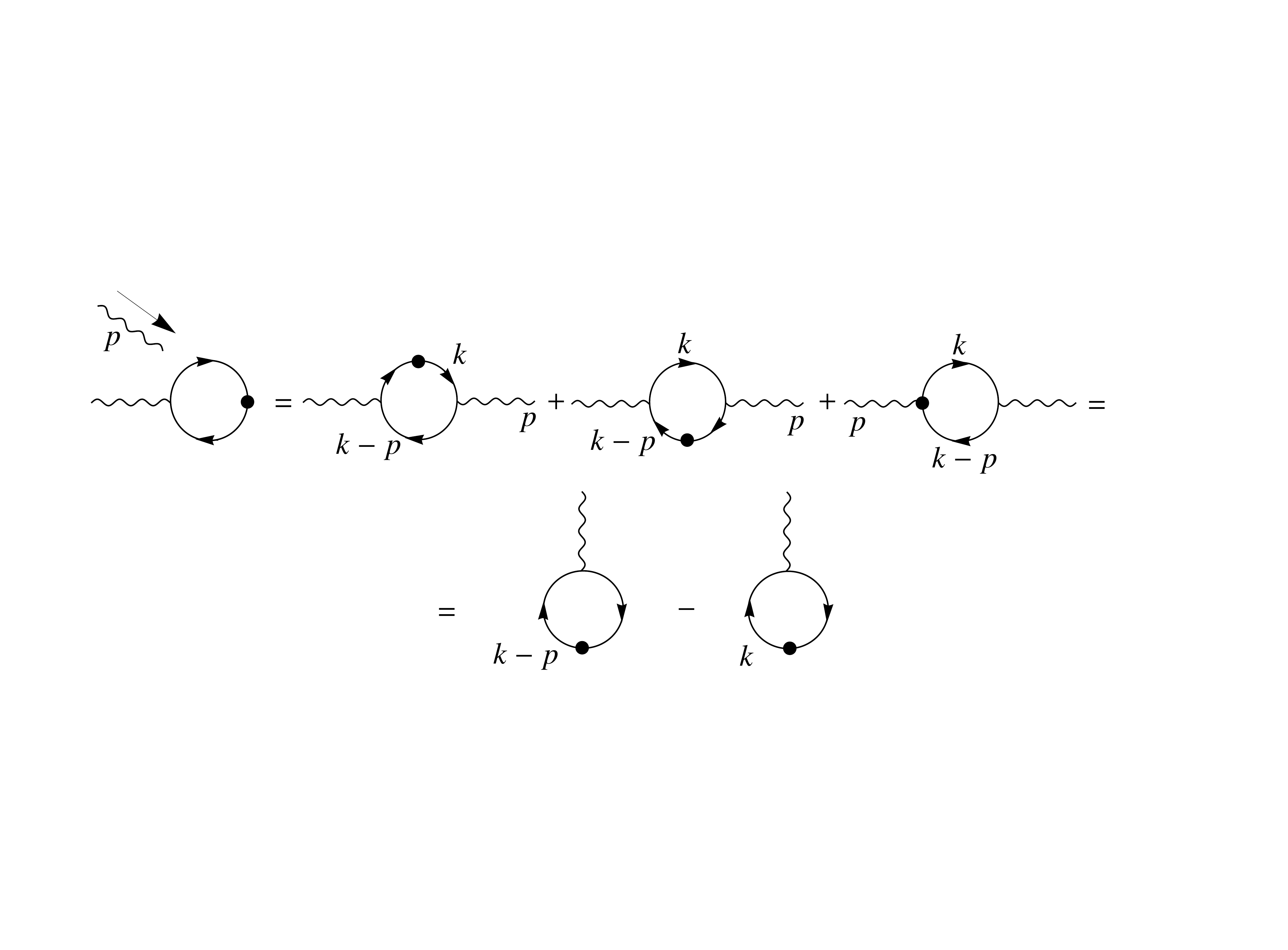}						
\caption{Relation between gauge and MRI.}     
\label{fig7}
\end{figure}

The main point to be noticed is that Ward identities can be depicted in diagrammatic form \cite{Peskin}. For example, to obtain the $n$-point function Ward 
identity, one needs to start with all $n-1$-point functions and perform an insertion of an external photon leg in all possible ways. In figure \ref{fig7}, we can 
see a simple example for two terms of the 2-point function Ward identity of the EQED.

Figure \ref{fig7} is a diagrammatic form of the simple following manipulation of the amplitudes:

\begin{align}
&p_{\mu}\Pi^{\mu\nu}(p)= \Big[\int_k Tr \left(\gamma^{\nu}\frac{1}{\slashed{k}-m}k_{\lambda}\Gamma_1^{\lambda}\frac{1}{\slashed{k}-m}\slashed{p}\frac{1}{\slashed
{k}-\slashed{p}-m}\right)\Big]+\Big[\int_k Tr \left(\gamma^{\nu}\frac{1}{\slashed{k}-m}\slashed{p}\frac{1}{\slashed{k}-\slashed{p}-m}(k-p)_{\lambda}\Gamma_1^{
\lambda}\frac{1}{\slashed{k}-\slashed{p}-m}\right)\Big]-
\nonumber\\&-\Big[\int_k Tr \left(\gamma^{\nu}\frac{1}{\slashed{k}-m}p_{\mu}\Gamma_1^{\mu}\frac{1}{\slashed{k}-\slashed{p}-m}\right)\Big]=\Big[\int_k Tr 
\left(\gamma^{\nu}\frac{1}{\slashed{k}-m}k_{\lambda}\Gamma_1^{\lambda}\frac{1}{\slashed{k}-\slashed{p}-m}\right)-\int_k 
Tr \left(\gamma^{\nu}\frac{1}{\slashed{k}-m}k_{\lambda}\Gamma_1^{\lambda}\frac{1}{\slashed{k}-m}\right)\Big]+\nonumber\\&+\Big[\int_k Tr \left(\gamma^{\nu}\frac
{1}{\slashed{k}-\slashed{p}-m}(k-p)_{\lambda}\Gamma_1^{\lambda}\frac{1}{\slashed{k}-\slashed{p}-m}\right)-\int_k Tr \left(\gamma^{\nu}\frac{1}{\slashed
{k}-m}(k-p)_{\lambda}\Gamma_1^{\lambda}\frac{1}{\slashed{k}-\slashed{p}-m}\right)\Big]
-\nonumber\\&-\int_k Tr \left(\gamma^{\nu}\frac{1}{\slashed{k}-m}p_{\mu}\Gamma_1^{\mu}\frac{1}{\slashed{k}-\slashed{p}-m}\right)=\int_k Tr \left(\gamma^{\nu}
\frac{1}{\slashed{k}-\slashed{p}-m}(k-p)_{\lambda}\Gamma_1^{\lambda}\frac{1}{\slashed{k}-\slashed{p}-m}\right)-\int_k Tr \left(\gamma^{\nu}\frac{1}{
\slashed{k}-m}k_{\lambda}\Gamma_1^{\lambda}\frac{1}{\slashed{k}-m}\right)
\label{Relation}
\end{align}

In equation (\ref{Relation}) we see that the 2-point function Ward identity is nothing but the difference between two tadpoles with different routing. Since the 
condition to preserve MRI is that the difference of the same Feynman diagram with two different routing is null, we easily see that by imposing MRI to the 
tadpoles depicted one automatically preserves gauge invariance. Even though the discussion above was restricted only to two of the six terms of the 2-point 
function Ward identity, the other terms can be shown to share the same behavior. Explicitly, one obtains by demanding MRI conditions:

\begin{figure}
\centering
 \includegraphics[clip, trim=20mm 150mm 70mm 130mm, scale=0.25]{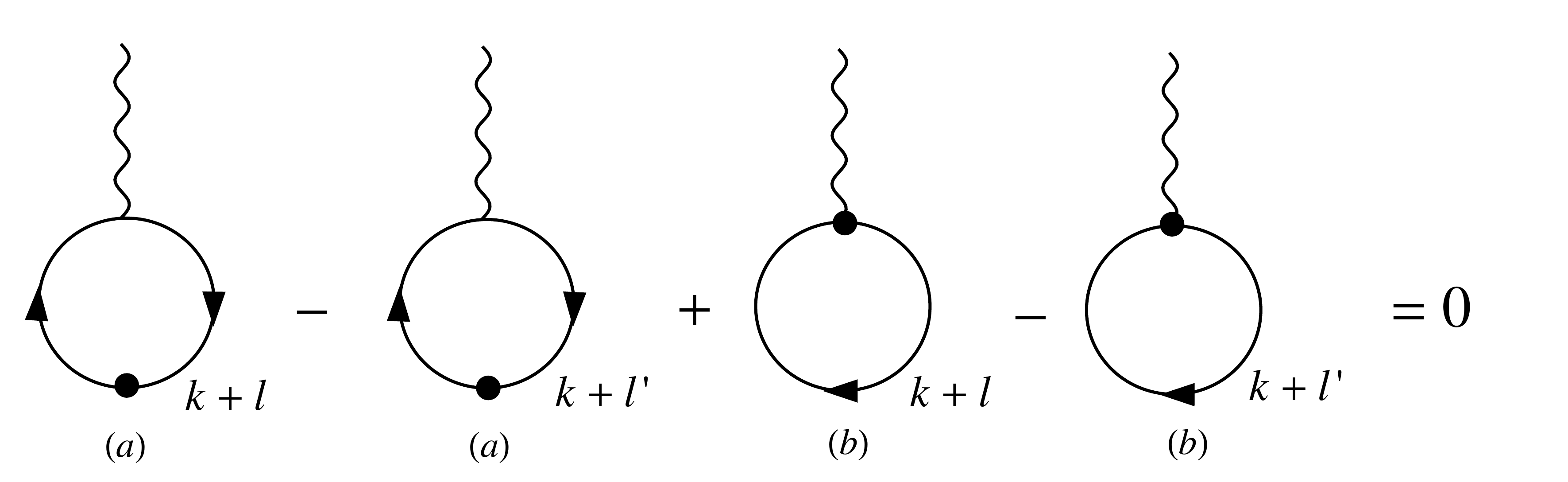}						
\caption{Momentum routing invariance of tadpoles.}     
\label{figMRI}
\end{figure}

In figure \ref{figMRI}, $l$ is an arbitrary routing that we can make proportional to the external momentum, {\it i.e.}, $l=\alpha p$. Let us call
the tadpoles a function of $l$, $\tau ^{\nu}(l)$. The result of figure \ref{figMRI} is

\begin{align}
&(\tau ^{\nu}_{(a)}(l)-\tau ^{\nu}_{(a)}(l'))+(\tau ^{\nu}_{(b)}(l)-\tau ^{\nu}_{(b)}(l'))=
4(\alpha'^3-\alpha^3)(12 c^{\alpha\beta}p_{\alpha}p_{\beta}p^{\nu}+4 p^{2}c^{\nu\alpha}p_{\alpha}+4p^{2}c^{\alpha\nu}p_{\alpha})(
\xi_0-\upsilon_0)+\nonumber\\&+4(\alpha^2-\alpha'^2)(p^2 m e^{\nu}+2 m p^{\nu}e\cdot p)(2\upsilon_0-\xi_0)+4(\alpha^3-\alpha'^3) (2c^{\alpha\beta}p_{\alpha}p_{\beta}p^{\nu}+c^{\nu\alpha}p_{\alpha}p^2+c^{\alpha\nu}p_{\alpha}p^2)\sigma_0+\nonumber\\&+4(\alpha'-\alpha)(c^{\nu\alpha}p_{\alpha}+c^{
\alpha\nu}p_{\alpha})(2\upsilon_2-\xi_2)
\label{Tad}
\end{align}

Since $\alpha\neq\alpha'$ by definition, the only possible solution of equation (\ref{Tad}) is to make all surface terms zero, {\it i.e.}, 
$\upsilon_0=\xi_0=\sigma_0=0$ and $\upsilon_2=\xi_2=0$. This solution assures gauge invariance since the Ward identity of equation (\ref{WIPI}) is
satisfied. 

\begin{figure}[!h]
 \includegraphics[trim=0mm 30mm 0mm 20mm,scale=0.55]{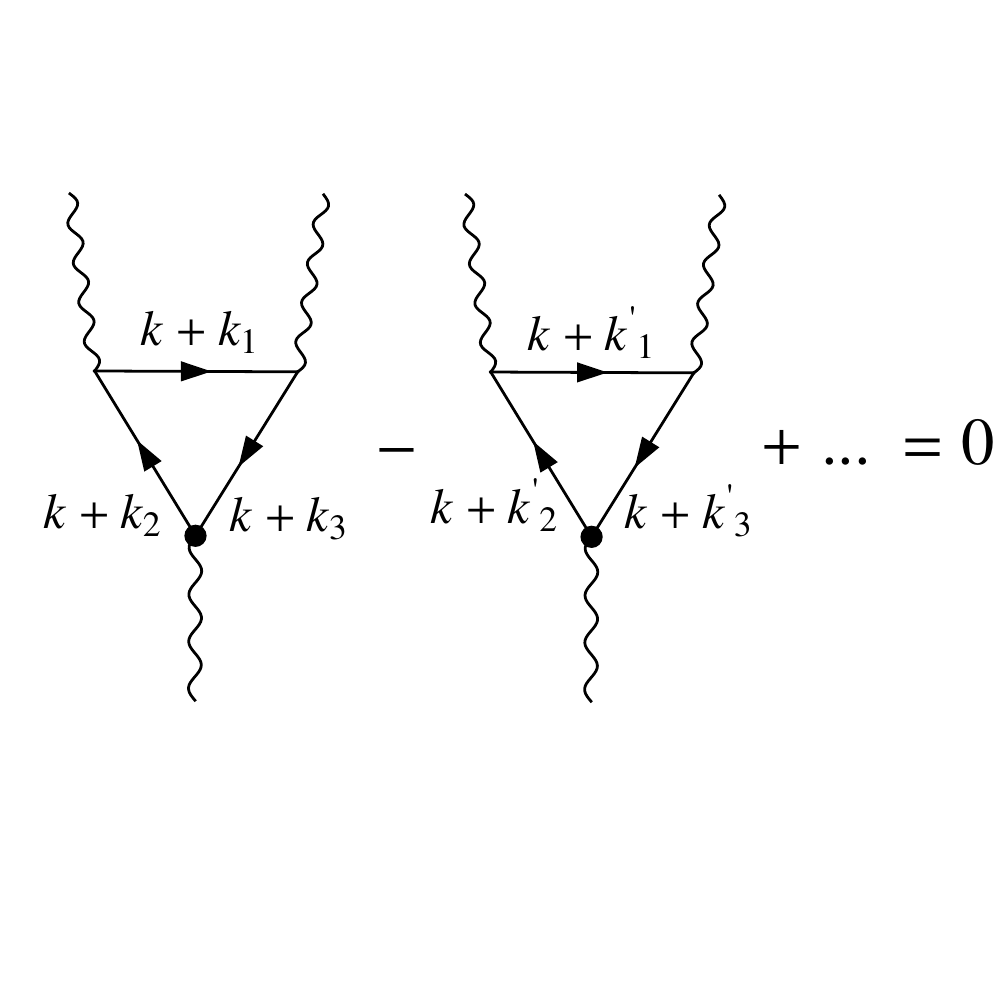}						
\caption{Diagrammatic representation of MRI.}     
\label{fig4}
\end{figure}

A similar feature occurs for the three-point functions as well. The MRI condition for the triangle diagrams is presented in figure
\ref{fig4}.

This time we have at least three different routings corresponding to each internal line. Thus, what we have to calculate is 

\begin{align}
 (p+q)_{\alpha}\sum_{i=a}^f(T^{\mu\nu\alpha}_{(i)}(k_1,k_2,k_3)-T^{\mu\nu\alpha}_{(i)}(k'_1,k'_2,k'_3))=0
 \label{EqMRI}
\end{align}

It is not necessary to consider diagrams $(g)$, $(h)$ and $(i)$ since they will contribute to the equation above with an $a_{\mu}$ coefficient that will not
affect the result or the conclusions bellow. 

The general routing $k_i$ obeys the following relations due to energy-momentum conservation at each vertex:
\begin{align}
k_2-k_3=& p+q,\nonumber\\
k_1-k_3=& p,\nonumber\\
k_2-k_1=& q.
\label{routing}
\end{align}

Equations (\ref{routing}) allow us to parametrize the routing  $k_i$ as combinations of external momenta:

\begin{align}
k_1=& \alpha p+(\beta-1) q,\nonumber\\
k_2=& \alpha p+\beta q,\nonumber\\
k_3=& (\alpha-1) p+(\beta-1) q.
\label{routing2}
\end{align}

Considering equations above, the result of equation (\ref{EqMRI}) implies that:
 
\begin{align}
 &(p+q)_{\alpha}\sum_{i=a}^f(T^{\mu\nu\alpha}_{(i)}(k_1,k_2,k_3)-T^{\mu\nu\alpha}_{(i)}(k'_1,k'_2,k'_3))= 8i(p+q)_{\alpha
}[(d_{\eta}^{\lambda}\epsilon^{\eta\mu\nu\alpha}+d_{\eta}^{\alpha}\epsilon^{\eta\nu\mu\lambda}+ d^{\nu}_{\eta}\epsilon^{\mu\eta\lambda\alpha}+d^{\mu}_{\eta}
\epsilon^{\nu\eta\alpha\lambda}+ \nonumber\\ &+d_{\kappa}^{\kappa}\epsilon^{\mu\nu\lambda\alpha})\upsilon_0-d_{\kappa}^{\kappa}\epsilon^{\mu\nu\lambda\alpha}\xi_0 
][(\alpha-\alpha')p_{\lambda}+(\beta- \beta')q_{\lambda}]=0
\label{tMRI}
\end{align}

Since $\alpha\neq\alpha'$ and $\beta\neq\beta'$ by definition, the only possible solution to the equation above is 
$\upsilon_0=\xi_0=0$. Finally, we conclude that MRI condition assures that gauge invariance holds at the quantum level, {\it i.e.}, 
$(p+q)_{\alpha}\sum_{i=a}^f  T^{\mu\nu\alpha}_{(i)}=0$, which means electric charge conservation in the SME beyond tree level.
That same conclusion was found in the usual QED \cite{Adriano}.

Of course, the same idea can be applied to an arbitrary number of external legs and loops. For instance, if the initial diagram has four (or more) external legs, 
as in figure \ref{fig7}, there is an equivalent equation which provides the difference between two diagrams with four (or more) external legs whose the 
internal labels are different. Since the finite part does not depend on the routing, gauge invariance is trivially respected in this case.

Besides, to make the complete one-loop proof we should calculate the Ward identity that comes from the Box diagrams. Although the calculation of the Box type 
diagrams are more involved from the computational viewpoint it can be straightforwardly shown that we would 
have an equivalent equation, as in figure \ref{fig7}, which would lead to the difference between two three leg diagrams with different routing. Since we show in 
equation (\ref{tMRI}) that these difference must be zero to assure the WI of the three point function, the WI of the Box diagrams must be zero as well.

\section{Concluding remarks}
\label{conclusion}

In this work we have studied the electromagnetic sector of the Standard Model Extension (SME) which encompasses all Lorentz-violating terms originated by 
spontaneous symmetry breaking at the Planck scale. In particular, we have verified under which conditions the above mentioned sector respects gauge invariance. 
We found, in consonant with the result of standard QED, that Ward identities are only broken by regularization-dependent objects. Since this objects only 
parametrize breakings by regularization methods themselves, we conclude that no quantum anomaly occurs in this model. We have also studied the interplay between 
MRI and gauge invariance and we have shown that they share a one-by-one correspondence even in the presence of chiral objects such as 
$\gamma_5$ matrix. This extends our previous proof for standard QED to a more broad class of models and allows us to conjecture MRI as a fundamental feature
that all theories should comply with. 

{\bf Acknowledgements}

The authors are grateful to professor V. Alan Kosteleck\'y for clarifying discussions and suggestions. A. R. V. acknowledges Brazilian financial agency CAPES, 
grant no. 99999.007290/2015-02, and IUCSS for the kind hospitality. A. L. C. acknowledges financial support by CNPq. M. S. acknowledges research grants from CNPq.

\section*{Appendix}
The results of all regularized integrals found after taking the trace in equation (\ref{WI1}) are

\begin{align}
\centering
&\int_k \frac{1}{[(k-p)^2-m^2]}= I_{quad}(m^2)-p^2\upsilon_0
\\
&\int_k \frac{k^{\alpha}}{[(k-p)^2-m^2]}= p^{\alpha}(I_{quad}(m^2)-\upsilon_2) -p^2p^{\alpha}(\xi_0-\upsilon_0)
\\
&\int_k \frac{k^{\alpha}}{[(k-p)^2-m^2]^2}= p^{\alpha}(I_{log}(m^2)-\upsilon_0)
\\
&\int_k \frac{k^{2}}{[(k-p)^2-m^2]^2}= I_{quad}(m^2)+(m^2+p^2)I_{log}(m^2) - 3 p^2 \upsilon_0
\\
&\int_k \frac{k^{\alpha}k^{\beta}}{[(k-p)^2-m^2]^2}= \frac{1}{2}g^{\alpha\beta}(I_{quad}(m^2)- \upsilon_2)+p^{\alpha}p^{\beta}(I_{log}(m^2)-\xi_0)
-\frac{1}{2}p^2 g^{\alpha\beta}(\xi_0-\upsilon_0)
\\
&\int_k \frac{k^{2}k^{\alpha}}{[(k-p)^2-m^2]^2}= 2 p^{\alpha}(I_{quad}(m^2)- \upsilon_2)+p^{\alpha}(m^2+p^2)I_{log}(m^2)+p^{\alpha}(3 p^2-m^2)\upsilon_0-
4 p^2 p^{\alpha}\xi_0
\\
&\int_k \frac{k^{\alpha}k^{\beta}k^{\gamma}}{[(k-p)^2-m^2]^2}= \frac{1}{2} p^{ \{ \alpha }g^{\gamma \beta \} }(I_{quad}(m^2)- \xi_2)+p^{\alpha
}p^{\beta}p^{\gamma}(I_{log}(m^2)+3\xi_0 - 3\upsilon_0 -\sigma_0)+\nonumber\\&+\frac{1}{2} p^2 p^{ \{ \alpha }g^{\gamma \beta \} }(2\xi_0 - \upsilon_0 -\sigma_0)
\\
\end{align}
The results of all regularized integrals found after taking the trace in equation (\ref{soma}) are

\begin{align}
\centering
&I=\int_k \frac{1}{(k^2-m^2)^2[(k+p)^2-m^2]}= -b\ \int^1_0 dx \frac{(1-x)}{\Delta^2}
\\
&I_1^{\beta}=\int_k \frac{k^{\beta}}{(k^2-m^2)^2[(k+p)^2-m^2]}= b\ p^{\beta}\ \int^1_0 dx \frac{x(1-x)}{\Delta^2}
\\
&J_1^{\beta}=\int_k \frac{k^{\beta}}{(k^2-m^2)[(k-p)^2-m^2]^2}= I_1^{\beta}+p^{\beta}I
\\
&I_2=\int_k \frac{k^2}{(k^2-m^2)^2[(k+p)^2-m^2]}= I_{log}(m^2)-b\ Z_0(p^2, m^2)-b\ m^2\ \int^1_0 dx \frac{(1-x)}{\Delta^2}
\\
&J_2=\int_k \frac{k^2}{(k^2-m^2)[(k-p)^2-m^2]^2}= I_2+p^2I+2p_{\beta}I_1^{\beta}
\\
&I_2^{\beta\nu}=\int_k \frac{k^{\beta}k^{\nu}}{(k^2-m^2)^2[(k+p)^2-m^2]}=\frac{1}{4}g^{\beta \nu}(I_{log}(m^2)-\upsilon_0 ) 
-\frac{1}{2} b\ g^{\beta \nu}[Z_0(p^2,m^2)-Z_1(p^2,m^2)]-\nonumber\\&-b\ p^{\beta}p^{\nu}\ \int^1_0 dx \frac{x^2(1-x)}{\Delta^2}
\\
&J_2^{\beta\nu}=\int_k \frac{k^{\beta}k^{\nu}}{(k^2-m^2)[(k-p)^2-m^2]^2}=I_2^{\beta\nu}+p^{\beta}p^{\nu}I+I_1^{\beta}p^{\nu}+I_1^{\nu}p^{
\beta}
\\
&I_3^{\nu}=\int_k \frac{k^2 k^{\nu}}{(k^2-m^2)^2[(k+p)^2-m^2]}= -\frac{1}{2} p^{\nu}(I_{log}(m^2)-\upsilon_0)+b\ p^{\nu}Z_1(p^2, m^2)+b\ m^2\ p^{\nu} \int^1_0 dx \frac{x(1-x)}{\Delta^2}
\\
&J_3^{\nu}=\int_k \frac{k^2 k^{\nu}}{(k^2-m^2)[(k-p)^2-m^2]^2}=I_3^{\nu}+p^{\nu}I_2+p^2 I_1^{\nu}+p^2 p^{\nu}I+ 2 p_{\gamma}I_2^{\gamma\nu}+ 2 p_{\gamma} p^{\nu 
}I_1^{\gamma}-p^{\nu}\upsilon_0
\\
&I_5^{\beta\nu\alpha}=\int_k \frac{k^{\beta}k^{\nu}k^{\alpha}}{(k^2-m^2)^2[(k+p)^2-m^2]}= \frac{-1}{12} b p^{\{ \alpha}g^{\beta \nu \}}(I_{log}(m^2)-\xi_0)+b\ p^{\alpha}p^{\beta}p^{\nu} \int^1_0 dx \frac{x^3(1-x)}{\Delta^2}+\nonumber\\
&+\frac{1}{2}b\ p^{\{ \alpha}g^{\beta \nu \}}[Z_1(p^2,m^2)-Z_2(p^2,m^2)] 
\\
&J_5^{\beta\nu\alpha}=\int_k \frac{k^{\beta}k^{\nu}k^{\alpha}}{(k^2-m^2)[(k-p)^2-m^2]^2}= I_5^{\beta\nu\alpha}+p^{\beta}p^{\nu}p^{\alpha
}I+p^{\{ \nu}p^{\beta
}I^{\alpha \} }_1 +p^{\{ \beta}I^{\nu\alpha \}}_2 +\frac{1}{4}p^{\{ \alpha}g^{\beta\nu\}}(\upsilon_0-\xi_0)
\end{align}
\begin{align}
\centering
&I_4^{\beta\nu}=\int_k \frac{k^2 k^{\beta}k^{\nu}}{(k^2-m^2)^2[(k+p)^2-m^2]}= \frac{1}{2}g^{\beta \nu}(I_{quad}(m^2)-\upsilon_2)+\frac{1}{4}(m^2-p^2)g^{\beta \nu}(I_{log}(m^2)-\upsilon_0)+\nonumber\\
&+\frac{1}{6}(p^2 g^{\nu \beta}+2 p^{\beta} p^{\nu})(I_{log}(m^2)-\xi_0)- b\ (-g^{\beta \nu}p^2+p^{\nu}p^{\beta})Z_2(p^2,m^2)+\frac{1}{2}\ b(m^2-3p^2)\ g^{\beta \nu} Z_1(p^2,m^2)+\nonumber\\
&+\frac{1}{2}b(p^2-m^2)g^{\beta \nu}Z_0(p^2,m^2)-b\ p^{\beta}p^{\nu} m^2\ \int^1_0 dx \frac{x^2(1-x)}{\Delta^2}
\\
&J_4^{\beta\nu}=\int_k \frac{k^2 k^{\beta}k^{\nu}}{(k^2-m^2)[(k-p)^2-m^2]^2}= \frac{1}{2}g^{\nu\beta}(I_{quad}(m^2)-\upsilon_2)+
(m^2-p^2)J_2^{\beta\nu}+2p_{\lambda}J_5^{\beta\nu\lambda}-2p_{\lambda}I_5^{\beta\nu\lambda}-p^2 I_2^{\beta\nu}
\end{align}

where $b\equiv \frac{i}{(4\pi)^2}$, $Z_k(p^2,m^2)$ and $\Delta^2$ are defined as
\bq
Z_k(p^2,m^2)=\int^1_0 dz z^k \ln \frac{m^2-p^2 z(1-z)}{m^2},\\
\Delta^2=m^2-p^2 x(1-x).
\eq

The basic divergent integrals $I_{log}(m^2)$ and $I_{quad}(m^2)$ and the surface terms $\upsilon_0$, $\upsilon_2$ and $\xi_0$ are
defined in section \ref{IReg}.

\end{document}